\documentstyle[pra,aps]{revtex}
\newcommand\be{\begin{equation}}
\newcommand\ee{\end{equation}}
\newcommand\bea{\begin{eqnarray}}
\newcommand\eea{\end{eqnarray}}

\begin{document}
\draft
\title{Quantum Parametric Resonance}
\author{Stefan Weigert\cite{thanks}} 
\address{Institut de Physique, Universit\'e de Neuch\^atel,
CH-Neuch\^atel\\
and\\
Department of Mathematics, University of Hull, UK-Hull\\
\tt s.weigert@hull.ac.uk}
\date{June 2001}
\maketitle
\begin{abstract}
The quantum mechanical equivalent of parametric resonance is studied. A simple model of a periodically kicked harmonic oscillator is introduced which can be solved exactly. Classically stable and unstable regions in parameter space are shown to correspond to Floquet operators with qualitatively different properties. Their eigenfunctions, which are calculated exactly, exhibit a transition: for parameter values with classically {\em stable} solutions the eigenstates are normalizable while they cannot be normalized for parameter values with classically {\em instable} solutions. Similarly, the spectrum of quasi energies undergoes a specific transition. These observations remain valid qualitatively for arbitrary linear systems exhibiting classically parametric resonance such as the paradigm example of a frequency modulated pendulum described by Mathieu's equation.
\end{abstract}
%
%
\newfont{\ggr}{cmsy8}
\newfont{\gkl}{cmmi8}
\section{Introduction}
The highly complicated behaviour of classically chaotic systems does not translate 
in an obvious and straightforward way into properties of their quantum counterparts. Nevertheless, various features such as energy level statistics and the spatial structures of wave functions have been identified \cite{reichl92,haake92} as more or less faithful 
quantum indicators of chaos in the classical limit. The extreme sensitivity of classically chaotic systems to the variation of initial
conditions appears, for fundamental reasons,  to figure less prominent in quantum systems.

Parametric resonance is a useful concept in order to understand the transition from regular to chaotic motion in classical dynamic systems. Each of the phase space filling tori 
of an integrable system (with two degrees of freedom) is characterized by a frequency ratio, and the value of this parameter determines the fate of a torus under a perturbation according to the KAM theorem \cite{lichtenberg+92}. When the strength of the perturbation increases, initially small areas of instability proliferate in the form of {\em Arnold tongues}.
They echo the intricate structure of phase-space regions where motion on tori and irregular trajectories coexist. Apart from its conceptual importance, parametric resonance has found applications in systems with many degrees of
freedom providing, for example, an important mechanism in the formation of
patterns \cite{manneville90,cross+93}. 

The purpose of this paper is to study the phenomenon of parametric resonance in a quantum context. Hamiltonians quadratic in
position and momentum are known to provide relevant examples of classical parametric
resonance. Since for such systems, classical and quantum mechanics of systems are intimately related, the program here is to make explicit 
what the {\em quantum manifestations} of the {\em classical instability}
look like. In particular, the focus will be on the quasi-energy spectrum of the quantum mechanical evolution operator or {\em Floquet} operator as well
on its eigenstates. It will be shown that the properties of the quantum system unambiguously reflect the stability and instability of its classical counterpart.

In Section \ref{defkho} classical parametric resonance is discussed from a general point of view. An analytically solvable model which exhibits parametric resonance on the classical level, is introduced in Section \ref{model}: a harmonic oscillator is subjected to a perturbation which periodically dilates and squeezes volumes in phase space. The parameter ranges for the instability of the classical system are determined. In Section \ref{qmpr}, Floquet eigenstates and quasi energies of the associated quantum system are calculated in the unstable case. By means of an {\em effective} Hamiltonian a comprehensive point of view for the possible scenarios is developped in Section \ref{effham}. Subsequently, the result is shown to persist for linear systems with arbitrary periodic frequency modulation. The generalization includes systems described classically by Mathieu's equation. Finally, the results are summarized and links to related models are established. 
\section{Classical parametric resonance}
\label{defkho}
Parametric {\em resonance} occurs if an appropriate parameter of a classical dynamical system is varied periodically in time. Stable fixed points of the flow in phase 
space become unstable for specific values of the period of the parameter 
variation. As an example, consider a pendulum with mass $m$ and 
length $l$ under the influence of gravity the support of which is moved up and 
down by an amount $\Delta l < l$ with frequency $\omega_0$. The equilibrum position of the 
undriven system corresponds to the bob resting vertically below the support. For small
oscillations, the pendulum is described by Mathieu's equation 
\begin{equation}
\frac{d^2 x}{dt^2} + \omega^2 (t) x = 0 \, ,   
\label{freqmod}
\end{equation}
with 
\begin{equation}
\omega^2 (t) = l -\Delta l \cos (\omega_0 t) > 0 \, ,   
\label{mathieufreq}
\end{equation}
where $x(t)$ is the vertical coordinate of the bob. This equation provides the paradigm for parametric resonance in classical mechanics: the behaviour of its solutions has been investigated in great detail \cite{abramowitz+84} as a function of the parameters $l, \Delta l$, and $\omega_0$ (see \cite{yorke78} for an interesting discussion).  
The equilibrium position is {\em destabilized} for appropriately chosen values of $\Delta l$ and $\omega_0$: arbitrarily small deviations from the fixed point will grow exponentially in time. This is characteristic for parametric resonance. For other parameter values the elliptic fixed point of the undriven system survives the perturbation. Smooth boundaries decompose the parameter space into regions with stable and unstable behaviour. On the separating boundary, a third marginal type of behaviour occurs which interpolates between stable and unstable motion.
 
Parametric excitation can also have the opposite effect which is less intuitive: an unstable fixed point may become a stable one. Indeed, the
{\em inverted} pendulum turns stable under specific conditions on $\Delta
l$ and $\omega (t)$ a phenomenon known as parametric {\em stabilization}. Small deviations from the vertical position do {\em not} grow without bound but the pendulum remains in the neighbourhood of its initial position for all times. 

Qualitatively, these phenomena show also up for other types of {\em periodic} driving,
$\omega^2 (t+T) = \omega^2(t)$, with a modified division of parameter space 
into stable and unstable regions. In the following, a particularly simple quantum system, known to exhibit parametric resonance classically, will be studied in detail. 
\section{A simple model}
\label{model}
Consider a harmonic oscillator with frequency $\omega$, $H_0 = p^{2}/2m 
+ m \omega^{2}x^{2}/2$, subjected to a periodic perturbation \cite{morath96},
\begin{equation}
H(t) = H_0 + H_k (t) \, ,
\label{classosc}
\end{equation}
where
\begin{equation}
H_k(t) 
= \frac{\alpha}{2}(xp+px) \delta_T (t) \, ,
      \qquad\alpha \in {\sf I} \! {\sf R} \, ,
\label{perper}
\end{equation}
with  an infinite comb of delta functions,
\be
\delta_T (t)= \sum_{n=-\infty}^{\infty}\delta(t-nT) \, .
\label{comb}
\ee
In between the times $T_{n}=nT$, $n\in \sf Z\!\!Z\rm$, the particle with mass $m$ 
moves {\em freely} in a quadratic potential, while at times $T_{n}$ it experiences  an impulsive kick with amplitude $\alpha xp$. In fact, the Hamiltonian does not depend on three but only two parameters ($\omega T$ and $\alpha$) as is seen from introducing $ \tau =t/T$, and rescaling simultaneously $p \to \sqrt{m\omega}p, q \to q/\sqrt{m\omega}$.

From a general point of view, the momentum dependence of the kick is a particular feature of this system. Other driven systems which have been studied as models for chaotic motion in both classical and quantum mechanics, such as the kicked rotator \cite{casati+79}, have a position dependent amplitude only. Various authors pointed out the consequences  which result from this difference \cite{casati+95,peres95}, and explicitly solvable models have been 
studied \cite{weigert93}.  

Consider the evolution of the system over one period, from 
$t=-\varepsilon$ to $ t= T-\varepsilon$, say, for $\varepsilon \ll T$. The equations of motion are
\begin{equation}
\frac{d}{dt} 
\left( \begin{array}{c} 
                            x \\
                            p
\end{array} \right)
= 
\left( \begin{array}{cc} 
                            \alpha \delta_T (t)  & 1/m\\
                            -m \omega^2 & -\alpha \delta_T (t)
                          \end{array} \right)
\left( \begin{array}{c} 
                            x \\
                            p
                          \end{array} \right) \, ,
\label{eqsmotion}
\ee
The integration over the time interval $-\varepsilon < t < \varepsilon$,  is carried out by first introducing a smooth approximation $\delta_T^\varepsilon (t)$ of the delta functions in (\ref{eqsmotion}). Then the solution of (\ref{eqsmotion}) is given by
\be 
\label{smoothdelta}
 \left( \begin{array}{c} 
                            x( +\varepsilon) \\
                            p( + \varepsilon)
\end{array} \right)
= 
{\cal T} \exp \left[ \int_{-\varepsilon}^{+\varepsilon} dt M_\varepsilon (t) \right] 
\left( \begin{array}{c} 
                            x( -\varepsilon) \\
                            p( - \varepsilon)
\end{array} \right) \, ,
\ee
where the symbol $\cal T$ denotes time ordering, and $M_\varepsilon (t) $ is the smooth $2 \times 2$ matrix from  (\ref{eqsmotion}). A straighforward calculation gives a simple result in the limit $\varepsilon \to 0$,
\be
\label{kickmap}
M_k 
\equiv \lim_{\varepsilon \to 0}  
\exp \left[ \int_{-\varepsilon}^{+\varepsilon} dt M_\varepsilon (t) \right]
= 
\exp \left( \begin{array}{cc} 
                            \alpha & 0\\
                            0 & -\alpha
                          \end{array} \right)
= 
\left( \begin{array}{cc} 
                            e^\alpha & 0\\
                            0 & e^{-\alpha}
                          \end{array} \right) \, .
\end{equation}
Consequently, the interaction $H_k(t)$ rescales instantaneously position and momentum by the amounts $\exp\lbrack\pm\alpha\rbrack$, respectively. In other words, it generates periodically {\em dilations} \cite{moshinsky+96},  which preserve volume in phase space. 

Next, Eqs. (\ref{eqsmotion}) are integrated over the remaining time interval between the kicks, $\varepsilon < t < T - \varepsilon$. In the limit $\varepsilon \to 0$, one obtains, 
\bea
\label{freemap}
M_0 &=& \lim_{\varepsilon \to 0} M_0^\varepsilon \nonumber \\
       &=&  \lim_{\varepsilon \to 0} \, \exp \left[ \left( 
                             \begin{array}{cc} 
                              0  & 1/m\\
                            -m \omega^2 & 0
                          \end{array} \right) (T-\varepsilon) \right]
      = \left(
           \begin{array}{cc}
              \cos \omega T  & (m\omega)^{-1} \sin \omega T \\
              -m\omega\sin \omega T & \cos \omega T
\end{array} \right) \, .
%
\eea

 Writing $z=(x,p)^{T}$ for points in phase space, the time evolution of the kicked harmoic oscillator (\ref{classosc}) over one full period from $t=0^{-} $ to $t= T^{-}$: $z(T^-) = M z(0^-)$,  is thus due to a kick followed by free motion, generated by the matrix 
\begin{equation}
M = M_0 M_k
= \left(
\begin{array}{cc}
 e^\alpha \cos(\omega T)        &  (e^{\alpha} m\omega)^{-1} \sin\omega T\\
-e^\alpha m\omega\sin\omega T &  e^{-\alpha} \cos \omega T
\end{array} 
\right) \, ,
\label{fullmap}
\end{equation}
being symplectic and of unit determinant. 

Dilations and oscillatory motion compete with each other. The overall character of the motion over one period depends on the actual values of the parameters. The matrix $M$ does not vary over phase space. Hence, for given values of 
the parameters $\omega$, $\alpha$ and $T$, all initial values $z \,  (\neq 0)$ evolve in a similar way. The matrix $M$ in (\ref{fullmap}) can generate three qualitatively different types of motion,  conveniently characterized by their eigenvalues 
\begin{eqnarray}
\lambda_\pm 
  &=& \frac{1}{2} \mbox{Tr } M \pm \sqrt{\frac{1}{4} (\mbox{Tr }M )^2 -1}                                                              \nonumber \\
  &=& \cosh \alpha \cos \omega T \pm\sqrt{\cosh^{2}\alpha \cos^{2}\omega T -1} \, .
\label{geneigenvalues}
\end{eqnarray}
The eigenvalues may come as
\begin{itemize} 
\item[$(\imath)$]  a complex conjugate pair with modulus one: $\lambda_\pm = \exp [\pm i\Omega]$; 
\item[$(\imath\imath)$]  a real reciprocal pair: $|\lambda_\pm | = \exp [\pm \mu]$; 
\item[$(\imath\imath\imath)$] a real degenerate pair with modulus one: $\lambda_+ = \lambda_- = \pm 1$. 
\end{itemize}
If the eigenvalues of $M$ are purely complex  ($\imath$), the images of a point of phase space $z$ under the composed action of $M_0$ and $M_k$ are all located on an {\em ellipse}. Appropriate rescaling of the axes transforms it into a circle such that during one period the angular coordinate of a point is seen to increase by an angle ${\widetilde \omega}$ defined through $\cos {\widetilde \omega} \equiv \cosh\alpha \cos \omega T$. The perturbation changes the original motion 
only {\em quantitatively}, {\em i.e.} the frequency of the oscillator now is ${\widetilde \omega}$ instead of $\omega$.  

In case ($\imath\imath$),  the real eigenvalues of $M$ indicate a ``hyperbolic rotation" parametrized conveniently by the positive real number $\mu$ with  
$\cosh \mu \equiv \cosh\alpha\cos\omega T $. Now the time evolution of the oscillator is changed {\em qualitatively} by the perturbation $H_k (t)$, from  stable to unstable motion. According to the sign of the eigenvalues $\lambda_\pm$ the iterates of a phase-space point are located on one (ordinary hyperbolic) or two (hyperbolic with reflection) branches of a {\em hyperbola}. The parameter regions associated with cases $(\imath)$ and 
$(\imath \imath)$ exhaust the parameter space almost completely. 

Stable and unstable regions are separated by boundaries defined by the condition $|\cosh\alpha \cos\omega T |=1 $. For the parameter values of case  ($\imath\imath\imath$), iterates of phase-space points are either on {\em one} ($\lambda_+ = \lambda_- = + 1$) or on {\em two} ($\lambda_+ = \lambda_- = - 1$) {\em straight lines}. This situation may be thought to interpolate between the elliptic and hyperbolic case. 

More precisely, the three cases are related with specific invariant curves in phase space. To see this, associate a quadratic form with the matrix $M$ by 
\begin{equation}
Q(z) = z^T \Lambda M z \, , \qquad 
\Lambda =  
\left(
\begin{array}{cc}
 0 & 1 \\
-1 & 0
\end{array} 
\right) \, . 
\label{quadraticform}
\end{equation}
In fact, $Q(z)$ is invariant under time evolution over one period:    
\begin{equation}
Q(M z) = Q(z) \, , 
\label{invariance}
\end{equation}
as follows from the symplecticity of $M$: $M^T \Lambda M = \Lambda$. 
Using $M_0$ and $M_k$ as examples for ($\imath$) 
and ($\imath\imath$), respectively, one obtains
\begin{equation}
Q_0(z) \propto \frac{1}{2}\left(p^2+ m^2 \omega^2 x^2 \right)\, , \qquad  
Q_k(z) \propto px \propto (p+x)^2 - (p-x)^2 \, ,
\label{explicitforms}
\end{equation}
while the marginal situation ($\imath\imath\imath$) implies  
\begin{equation}
Q_{\imath \imath \imath} (z) \propto (p\pm x)^2 \, ,  
\label{explicitformmarg}
\end{equation}
using $ \cos \omega T = \pm 1/ \cosh \alpha$ and $\sin \omega T = \tanh \alpha$. These results can be summarized by writing 
\be
\label{qformsinone}
Q (z) \propto \frac{1}{2} \left( P^2 + \Omega^2 X^2 \right) \, , 
\ee
where $(X,P)$ are related to $(x,p)$ by appropriate linear canonical transformations.
Then, a variation of the original parameters is reflected in a change of the factor $\Omega^2$ multiplying the quadratic term in $X$: $\Omega$ is either real, purely imaginary, or zero. In physical terms, the time evolution of the system over one period is effectively that of a particle in an attractive or repulsive quadratic potential, or in the absence of a potential. Not surprisingly, the quadratic forms $Q(z)$ will play an important role for  quantum parametric resonance.
 
If $\cos\omega T=1$, or $\omega T=2\pi k$, $k\in \sf Z\!\!Z\rm$, an arbitrarily small kick amplitude $\alpha$ renders the system unstable. 
This particularly simple situation---the harmonic evolution has{\em no} net  effect: $M_0 = 1$,---is called {\em resonant}.  Similarly, for $\omega T=2\pi(k+1/2)$, $k\in\sf Z\!\!Z\rm$, the time evolution is {\em resonant with reflection}: each point $z$ in phase space is mapped to $(-z)$ by the harmonic time evolution: $M_0= -1$.

\section{Quantum parametric resonance}
\label{qmpr}
The quantum mechanical harmonic oscillator with an impulsive force \cite{morath96}
 is described by the Hamiltonian operator 
\begin{equation}
\hat {H}(t)
    = \frac{\hat{p}^{2}}{2m}+\frac{m\omega^{2}}{2}\hat{x}^2
      + \frac{\alpha}{2}(\hat{x}\hat{p}
      + \hat{p}\hat{x}) \delta_T (t) 
%
\label{qmoscill}
\end{equation}
Position and momentum operators $\hat{x}$ and $\hat{p}$ satisfy the fundamental commutation relation $\left[\hat{x},\hat{p}\right]=i\hbar$. The long-time behaviour of the system is determined by the Floquet operator ${\cal F} = U(t_0 + T,t_0) \equiv U_T$. It maps a state at time $t_{0}$ over one period to another state: ${\cal F}|\psi(t_0)\rangle = |\psi(t_0+T)\rangle$. 
Choose the time $t_0 = T_0^- = 0^-$ just before the kick at $t=0$. Then, due to the $\delta$-type interaction the Floquet operator is a product of two unitary operators
\begin{equation}
{\cal F} =  U_0 U_k \, .
\label{floquet}
\end{equation}
The first unitary generates harmonic motion with frequency $\omega$ from $ t=0^+$ to $t=T_-$:  
\begin{equation}
U_0(T^{-},0^{+})
   = {\cal T} \int_{0^{+}}^{T^{-}}dt\exp\left[-i\hat{H}(t)/\hbar\right]
   = \exp\left[-\frac{i}{\hbar} \left( \frac{\hat{p}^{2}}{2m}
     + \frac{m\omega^{2}\hat{x}^2}{2} \right) T \right] \, ,
\label{unitaryfree}
\end{equation}
where ${\cal T}$ denotes time ordering again. The second unitary describes the effect of the kick at time $t=0$. As before, the time-ordered product is evaluated explicitly  by approximating the $\delta$ distribution as a strongly peaked, smooth function $\delta_T^\varepsilon$ and taking the limit $\varepsilon \to 0$ \cite{weigert93},
\begin{equation}
U_k(0^{+},0^{-}) = {\cal T} \int_{0^{-}}^{0^{+}}dt\exp\left[-i\hat {H}(t)/\hbar\right]
      = \exp\left[-\frac{i\alpha}{2\hbar}(\hat{x}\hat{p}+\hat{p}\hat{x})\right] \, .
\label{unitarykick}
\end{equation}
As shown in \cite{facchi+00}, the resulting ``squeeze operator'' \cite{stoler70} has an implementation in a quantum optical context. Some properties of the operator in the exponential in (\ref{unitarykick}) have been studied in \cite{bollini+93}.    
Using the shorthand $\hat z \equiv (\hat x, \hat p)^T$, the action of the Floquet operator ${\cal F}$ on position and momentum operators works out as in the classical case:
\begin{equation}
\hat z(T^{-}) = {\cal F} \hat z(0^{-}) {\cal F}^\dagger  
              = M_0 M_k \hat z(0^{-})
              = M \hat z(0^{-}) \, ,
\label{heisenberg}
\end{equation}
where
\begin{eqnarray}
\hat z(T^-)   &=& U_0 \hat z(0^{+}) U_0^\dagger = M_0 z(0^{+}) \, ,  \\
\hat z(0^{+}) &=& U_k \hat z(0^{-}) U_k^\dagger = M_k z(0^{-}) \, . 
\label{qmevol}
\end{eqnarray}
These relations follow from either integrating the linear Heisenberg equations of motion or from expanding and resumming the exponentials involved. On comparing Eqs. (\ref{kickmap},\ref{freemap}) to the result (\ref{heisenberg}), the intimate relation between the classical and quantum time evolutions generated by a quadratic Hamiltonian is evident.

Therefore, the quantum system is expected to inherit the division of parameter space into stable and unstable regions on the basis of Eqs. (\ref{heisenberg}), {\em i.e.} through the properties of the matrix $M$. How do the different parameter regions manifest themselves in the quantal framework? 
When moving in parameter space from a classically stable to an unstable region, the eigenstates $|\phi_{\mu}\rangle$ as well as the quasienergies $E_{\mu}$ of the Floquet operator,
\begin{equation}
{\cal F} |\phi_{\mu}\rangle 
   = \exp(-iE_{\mu}T/\hbar)|\phi_{\mu}\rangle \, ,
\label{floqueteigeneq}
\end{equation}
will undergo qualitative changes. The merit of the present model is that one can determine $|\phi_{\mu}\rangle$ and $E_{\mu}$ explicitly 
for characteristic cases. This will be done in Section 5 by means of an 
{\em effective} Hamiltonian $\hat H_{\mbox{\footnotesize eff}}$ which generates the time evolution over one period. 

Before presenting the unified treatment, the most interesting case of a Floquet operator associated with a classically {\em unstable} region of parameters will be studied in detail. Consider, for simplicity, the resonant case, defined by $\omega T = 2\pi k, k  \in \sf Z\!\!Z\rm $:  the classical motion becomes unstable for any nonzero value of $\alpha$. The Floquet operator ${\cal F}$ reduces to just the kick operator $U_k$. It is the simplicity of the action of the kick on position and momentum eigenstates which allows one to construct the complete set of ($\delta$-normalized) eigenstates of the time evolution operator.

According to (\ref{qmevol}), the position operators $\hat x_\pm $ just before and after the kick at time $t=0$ are related by
\be
\label{beforeafter}
\hat x_+ = U_k \hat x_- U_k^\dagger = e^\alpha \hat x_- \, ,
\ee
and the eigenvalue equations read
\begin{equation}
\hat x_+ |x_+ \rangle = x_+ |x_+ \rangle \, ,
\qquad \hat x_- |x_- \rangle = x_-|x_- \rangle \, .
\label{position}
\end{equation}
Multiplication of the second equation with $U_k$ from the left and using (\ref{beforeafter}) implies that 
\be 
\label{close}
\hat x_- \left( U_k |x_- \rangle \right)  
 = e^{-\alpha} x_-\left( U_k |x_- \rangle \right) \, .
\ee
Hence, the state $ U_k |x_- \rangle $ is an eigenstate of $\hat x_-$; in other words, the operator $U_k$ maps a position eigenstate $|x\rangle$ to another position eigenstate with eigenvalue $e^{-\alpha}x$:
\begin{equation}
 U_k|x\rangle = \frac{1}{\sqrt{\rho(x)}} |e^{-\alpha} x \rangle \, .
\end{equation}
The factor $1/\sqrt{\rho(x)}$ accounts for a possible change of the normalization of the states. It guarantees the completeness relation to hold in the form $\int_{-\infty}^{\infty}dx|x\rangle\langle x|=1$ (cf. \cite{bohm93}), and it is fixed by the following condition: 
\begin{eqnarray}
\delta(x-x') &=& \langle x|x'\rangle 
              =  \langle x| U^{\dagger}_k  U_k |x'\rangle 
              =  \frac{1}{\rho(x)} \langle e^{-\alpha}x | e^{-\alpha} x' \rangle \nonumber \\
             &=& \frac{1}{\rho(x)} \delta (e^{-\alpha}(x-x'))
              =  \frac{e^{\alpha}}{\rho(x)}\delta(x-x') \, .
\label{normalize}
\end{eqnarray}
This requires $\rho(x)=e^{\alpha}$ leading to
\begin{equation}
U_k|x\rangle = e^{-\alpha/2}|e^{-\alpha}x\rangle \, ,
\label{includenormx}
\end{equation}
which agrees with the result in \cite{casati+89}. The action of $U_k$ on a momentum eigenfunction $|p\rangle$ reads
\begin{equation}
U_k|p\rangle = e^{\alpha/2}|e^{\alpha}p\rangle \, ,
\label{includenormp}
\end{equation}
as follows from  a similar argument for $\hat p$ or from exploiting   $\langle p | x \rangle = \langle p | U_k^\dagger U_k | x \rangle$, etc. 

The transformation property (\ref{includenormx}) provides the clue to determine the Floquet eigenstates as linear combinations of position eigenstates. Imagine to iterate a position 
eigenstate $|x_{0}\rangle$ forwards and backwards by applying powers of $U_k$ and $U_k^{-1} = U_k^\dagger$, respectively. After multiplying with appropriate phase factors candidates for eigenstates of $ {\cal F} \equiv U_k $ have the form:
\begin{equation}
|x_{0},\mu\rangle 
= \frac{1}{\sqrt{2\pi}}\sum_{n=-\infty}^{\infty}e^{i\mu n}
                               e^{-\alpha n/2}|e^{-\alpha n}x_{0}\rangle \, .
\label{feigenstate}
\end{equation}
As it stands, $|x_{0},\mu\rangle$ is indeed an (improper) eigenstate of $\cal F$ with eigenvalue $\exp [-i\mu]$ since
\begin{eqnarray}
{\cal F} |x_{0},\mu\rangle 
    &=& \frac{1}{\sqrt{2\pi}} \sum_{n=-\infty}^{\infty}e^{i\mu n}e^{-\alpha                         (n+1)/2}|e^{-\alpha (n+1)}x_{0}
          \rangle \nonumber \\
    &=& e^{-i\mu}\frac{1}{\sqrt{2\pi}}\sum_{n=-\infty}^{\infty}
         e^{i\mu n}e^{-\alpha n/2} |e^{-\alpha n} x_{0}\rangle \nonumber \\
    &=&  e^{-i\mu}|x_{0},\mu \rangle \, .
\label{checkeigenstate}
\end{eqnarray}
There is no restriction on the values of $\mu \in [0,2\pi)$--hence the spectrum of quasi energies is {\em continuous}. As one can check explicitly, the states $|x_{0},\mu\rangle$ do provide an orthonormal complete set in Hilbert space  
if the label $x_{0}$ is from either one of the two intervals $[1,e^{\alpha})$ or $[-e^{\alpha}, -1)$, $\alpha > 0$, say,
\begin{eqnarray}
\langle x_{0},\mu|x_{0}',\mu'\rangle
    &=& \frac{1}{2\pi} \sum_{m,n=-\infty}^{\infty}e^{-i\mu n}e^{i\mu' m}
         e^{-\alpha (n+m)/2} \langle x_{0}|(U_k^{\dagger})^{n} U_k^{m}|x_{0}'\rangle  \nonumber \\
    &=& \frac{1}{2\pi}\sum_{n,m=-\infty}^{\infty}
        e^{-i(\mu n - i\mu' m)} e^{-\alpha (n+m)/2} 
           \langle e^{-\alpha n}x_{0}|e^{-\alpha m} x_{0}' \rangle \delta_{nm} \nonumber\\
    &=& \frac{1}{2\pi}\sum_{n =-\infty}^{\infty} 
           e^{i(\mu'-\mu)n} e^{-\alpha n} 
           \delta \left( e^{-\alpha n} (x_{0} - x_{0}')\right) \nonumber \\                                                 
    &=& \delta(\mu -\mu')\delta(x_{0}-x_{0}') \, .
\label{orthon}
\end{eqnarray}
Note that for $x_{0}$ and $x_{0}'$ chosen from the {\em same} the reference interval, the scalar product 
of position iterates with {\em different} $n$ and $m$ vanishes and, thus, gives rise to a Kronecker delta $\delta_{nm}$. 

Contrary to one's first expectation, the continuous label $x_0 \in [1,e^\alpha]$ does not imply a continuous degeneracy of the quasi energies $E_\gamma$ but only a {\em countable} one. This is due to the fact that  eigenstates of the position operator $\hat x$ are improper ones, {\bf i.e.} not normalizable. To see this in a simple way, consider a particle in a 
one-dimensional box. Its energy eigenstates span a Hilbert space with countable dimension. Nevertheless, one needs continuously many states to span this space by (improper) states $|x\rangle$. In Section 5.3 the countable degeneracy of the quasi energies $E_\gamma$ will also follow from general considerations. 

The completeness relation reads
\begin{equation}
\int\limits_{0}^{2\pi}d\mu\int\limits_{1}^{e^{\alpha}}dx |x,\mu\rangle\langle x,\mu|
     + \int\limits_{0}^{2\pi}d\mu\int\limits_{-1}^{-e^{\alpha}}dx |x,\mu\rangle\langle x,\mu|
     + \lim_{\epsilon\to 0}\int\limits_{-\epsilon}^{+\epsilon}dx|x\rangle\langle x|
     = 1
\label{complete}
\end{equation} 
where the integration extends over both fundamental intervals, and the state $| x = 0 \rangle $ has to be included as an additional eigenstate of the Floquet operator. This state is clearly orthogonal to all other eigenstates of ${\cal F}$ in Eq. (\ref{feigenstate}).

The Floquet operator associated with a classically instable region is thus seen to 
have a continuous spectrum of quasi energies, and the supprot of its eigenfunctions in configuration space is not compact but extends to infinity. In addition, there is one state localized
at the origin, $x=0$, thus ``sitting on top'' of the instable fix point of the classical map.  
\section{Effective Hamiltonian}
\label{effham}
According to Eq.\ (\ref{floquet}) the Floquet operator ${\cal F}$ is a product of two noncommuting operators $U_0$ and $U_k$ which are quadratic functions of position $\hat x$ and momentum $\hat p$. Since the Floquet operator is unitary, on can express it in the form ${\cal F} = \exp [ -i \hat H_{\mbox{\footnotesize eff}} T /\hbar ]$, with an ``effective'' Hamiltonian $\hat H_{\mbox{\footnotesize eff}}$. This operator is obtained from entangling the two unitaries into a {\em single} exponential using Baker-Campbell-Hausdorff technology as presented in \cite{wilcox67},
\begin{equation}
%
\exp\left[-\frac{i}{\hbar} \left( \frac{\hat{p}^{2}}{2m}
     + \frac{m\omega^{2}\hat{x}^2}{2} \right) T \right]
     \exp \left[- \frac{i \alpha}{2 \hbar}\left(\hat x\hat p+\hat p\hat x \right) \right] \equiv \exp \left[- \frac{i}{\hbar} \hat H_{\mbox{\footnotesize eff}} T \right] \, .
\label{product}
\end{equation} 
In the present case, it is possible to determine explicitly the effective Hamiltonian since the operators in the exponents of the product in (\ref{product}) constitute a  closed Lie algebra,
\begin{eqnarray}
\left[\hat x^{2},\hat p^{2}\right]
   &=& 2 i \hbar \left(\hat x\hat p+\hat p\hat x\right) \, ,  \label{c1}\\
\left[\left(\hat x\hat p+\hat p\hat x\right), \hat p^{2} \right]
   &=& 2 i \hbar \hat p^{2} \, , \label{c2} \\
\left[\left(\hat x\hat p+\hat p\hat x\right), \hat x^{2}\right]
   &=& - 2 i \hbar \hat x^{2} \, ,
 \label{c3}
\label{splitalgebra}
\end{eqnarray}
, equivalent to the ``split three-dimensional algebra'' \cite{wilcox67}. Physical realizations of this algebra can be given in terms of creation and annihilation operators of a two-dimensional harmonic oscillator or angular momentum operators. 

Due to the commutation relations, the effective Hamiltonian is necessarily a linear combination of the three quadratic generators, 
\begin{equation}
- \frac{i}{\hbar} \hat H_{\mbox{\footnotesize eff}}T 
     = a \hat p^{2} + b \hat x^{2} + \frac{c}{2} (\hat x\hat p + \hat p\hat x) \, .
\label{linearcombi}
\end{equation}
The transformation of the operators $\hat p$ and $\hat x$ over one period $T$ through ${\cal F}$ has been determined in Eq. (\ref{heisenberg}), while in \cite{wilcox67} their transformation has been calculated starting from a {\em given} quadratic expression (\ref{linearcombi}). Combining these results, it is straightforward to determine the effective Hamiltonian: the parameters $a$, $b$, $c$ must be expressed in terms of the elements $M_{jk}$ $(j,k=1,2)$ of the matrix $M$. After some algebra one finds
\begin{eqnarray}
a &=& \frac{i}{\hbar} \frac{\Delta}{2} M_{12} 
\, , \\
b &=& - \frac{i}{\hbar}  \frac{\Delta}{2} M_{21}
\, , \\
c &=&  \frac{i}{\hbar}  \frac{\Delta}{2} (M_{11}-M_{22}) 
  \, .
\label{coefficients}
\end{eqnarray}
The real number $\Delta$ is given by 
\begin{equation}
\Delta
  = \frac{\mbox{arcsinh}D}{D} \equiv \Delta( D^2) \in \sf I \! R \rm \, 
\label{delta}
\end{equation}
being thus an even function of the difference between the eigenvalues of the matrix $M$,
\begin{equation}
D = \frac{1}{2}(\lambda_+ -\lambda_-) 
  =  \left( \frac{1}{4} ( M_{11}+M_{22})^{2}-1 \right)^{1/2} \, .
\label{diskr}
\end{equation}
Using (\ref{fullmap}) for the matrix elements $M_{jk}$, one obtains the expression 
\be
\hat H_{\mbox{\footnotesize eff}} 
= 
\Delta \frac{\sin \omega T}{\omega T} 
           \left( \frac{\hat p^2}{2m e^\alpha} 
                  +   \frac{m  \omega^2 e^\alpha}{2} \hat x^2 
                      + \omega \sinh \alpha \cot \omega T (\hat x \hat p + \hat p \hat x) \right) \, .        
\label{heffexp}
\ee
For $\alpha = 0$ or $\omega=0$ the relations above reproduce correctly the operators $U_0$ and $U_k$, respectively.  It is interesting to note that this result is closely related to the {\em quantized quadratic form}, i.e. the expression $\hat  Q = Q(\hat z)$ obtained from $Q(z)$ in (\ref{quadraticform}) by replacing $z \to \hat z$ and appropriately symmetrizing,
\be
Q(\hat z) 
\equiv \hat z^T \Lambda M \hat z  
= \sigma \hat H_{\mbox{\footnotesize eff}} + \tau \hat I \, ,
\label{quadraticinvariantandeffham}
\ee
with two constants $\sigma$ and $\tau$. Retrospectively, it is natural to find that the generator for time displacement over one period is a function of the quantized quadratic form invariant under the corresponding classical motion. 
In other words, the Floquet operator $\cal F$ commutes with the corresponding quadratic form.

Therefore, three types of effective quantum Hamiltonians will show up which are in correspondence with the classically stable, unstable, and marginal case. Each of them is unitarily equivalent 
to one of the following expressions:
\begin{equation}
\hat H_{\mbox{\footnotesize eff}} 
        \propto  \frac{1}{2} \left( \hat P^{2} + \Omega^{2} \hat X^{2} \right) \, ,
\label{unifiedeffham}
\end{equation}
where the operators $\hat P$ and $\hat X$ are unitarily 
equivalent to $\hat x$ and $\hat p$,  and there are three possible values for the frequency $\Omega$, nonzero real, purely complex or equal to zero. For simplicity, the discussion to follow will take (\ref{unifiedeffham}) as a starting point. There is no need to use the explicit expressions of $\hat H_{\mbox{\footnotesize eff}}$ in terms of the original variables. 
\subsection{Stable elliptic case}
For parameter values associated with a classically stable region, the effective Hamiltonian operator $\hat H_{\mbox{\footnotesize eff}}$ is unitarily equivalent to a harmonic oscillator with unit mass and real non-zero frequency $\Omega$. Consequently, the eigenstates of the Floquet operator satisfy:  
\begin{equation}
{\cal F} |n \rangle = e^{-i \varepsilon_n } |n \rangle \, , \qquad
           \varepsilon_n = E_n T/\hbar            \, ,
\label{ellipticsolutions}
\end{equation}
with  the familiar oscillator eigenstates $|n \rangle$,
\begin{equation}
\hat  H_{\mbox{\footnotesize eff}}|n \rangle 
     = E_n  |n \rangle \, , \qquad 
       E_n = \hbar\Omega \left( n+\frac{1}{2} \right) \, , \quad 
         n \in {\sf I \! N}_0 \, ,
\label{famoscill}
\end{equation}
which provide an orthonormal basis in Hilbert space. The spectrum of quasi energies $\varepsilon_n$ associated with the 
operator $\cal F$ in (\ref{ellipticsolutions}),
\begin{equation}
\varepsilon_{n} =  \Omega T
\left( n+\frac{1}{2} \right) \mbox{mod} \,  2\pi \, , \qquad    n = 0,1,2, \ldots ,  
\end{equation}
is obtained by `projecting' the oscillator spectrum $E_n$ onto the interval $[ 0 , 2 \pi )$. Two possibilities arise:
\begin{enumerate}
\item 
If $\Omega T $ is an {\em irrational} multiple of $2 \pi$ then all quasi energies $\varepsilon_{n}$ are different, $\varepsilon_{n} \not= \varepsilon_{n'}$. The union of all $\varepsilon_{n}$ is dense in the interval from $0$ to $2\pi$. The set of eigenvalues $\{\varepsilon_n \}$ is countable, and no quasi energy is degenerate. 
\item
If  $\Omega T$ is a {\em rational} multiple of $2 \pi$, say $r/s$,
then $\varepsilon_{n+s}=\varepsilon_{n}$ for all $n$, and there are only $s$ different quasi energies: $\varepsilon_{0}$,$\varepsilon_{1},\ldots,\varepsilon_{s-1}
\in [0,2\pi )$. Each value is countably often  degenerate, and the Hilbertspace $\cal H$ decomposes in a direct sum, 
\begin{equation}
{\cal H} = \bigoplus_{\nu=0}^{s-1} {\cal H}_{\nu} \, ,
\label{ellsubspaces}
\end{equation}
where the finite-dimensional degenerate subspace ${\cal H}_{\nu}$ is spanned by the states $\{ |\nu + k s\rangle, k \in {\sf I} \! {\sf N}_0 \} $.
\end{enumerate}
\subsection{Marginal case}
Consider now parameter values which imply that the effective Hamiltonian operator is unitarily equivalent to 
that one of a free particle: $\hat  H_{\mbox{\footnotesize eff}} \propto {\hat  P}^2 $, i.e. $\Omega = 0$. Its eigenstates are identical to those of a momentum-type operator, 
\begin{equation}
\hat  P |P \rangle 
     = P  |P \rangle \, , \qquad 
       P \in ( -\infty, \infty) \, ,
\label{marginalsolutions}
\end{equation}
hence require a $\delta$-normalization. 
Since these states are also eigenstates of the Floquet operator $\cal F$, it has a {\em continuous} set of $\delta$-normalized eigenstates. In the position representation the expansion coefficients of the states $| P \rangle$ have modulus one throughout, and the states do {\em not} decay for $X \to \pm \infty$. 

The quasi energies $\varepsilon_{0,k \pm } = P^2 /(2\hbar) $ take all values in the interval $[0, 2 \pi)$. Each quasi energy is countably often degenerate: given a state $| P_0 \rangle$, all states $| P_{0,k_\pm} \rangle$ with
\begin{equation}
P_{0,k}^\pm = \pm \sqrt{ P_0^2 
             + \frac{2\hbar}{T} 2\pi k } \, ,    
                   \qquad k = 1, 2, \ldots ,
\label{otherPs}
\end{equation}
are eigenstates with the same eigenvalue: $ \exp [- i (P_{0,k}^\pm)^2 T/(2\hbar)] = \exp [- i P_0^2 T/(2\hbar)] $. In analogy with (\ref{ellsubspaces}), the Hilbert space can be thought of as a continuous product of subspaces each 
with countable dimension.
\subsection{Unstable hyperbolic case} 
The quantum system associated with a classically instable region is unitarily equivalent to a particle in an inverted  
quadratic potential, that is, the Hamiltonian in (\ref{unifiedeffham}) with $\Omega^2 < 0$. The potential takes arbitrarily large negative values for $X \to \pm \infty$, hence the solutions oscillate ever faster for increasing $X$. No normalizable eigenfunctions exist but there are two solutions 
of Schr\"odinger's equation for every value of the energy $E$.  The spectrum of quasi energies $E_\mu$ exhibits thus the same features as in the marginal case: the numbers $E_\mu$ take any value between $0$ and $2 \pi$, and each value is countably often degenerate. This is due ot the exponential function which ``wraps'' the real variable $E \in (-\infty , \infty)$ around a circle. For the resonant case, the solutions have been given explicitly in Section 4 as a sum of delta functions in the position representation with nonuniform amplitudes.
\section{Arbitrary frequency modulation}
\label{general}
The results obtained so far have been derived for a model with a particularly simple time evolution. Classically, however, the phenomenon of parametric resonance is known to occur for much more general periodic frequency modulations. In fact, the behaviour of the associated quantum systems can be shown to exhibit qualitatively the same behaviour.  Consider the classical Hamiltonian 
\begin{equation}
H(t) = \frac{p^{2}}{2m} + \frac{m \omega^{2}(t)}{2}x^{2} \, , \qquad \omega^{2}(t+T) = \omega^{2}(t) \, ,
\label{freqmodham}
\end{equation}
which generates the equations of motion (\ref{freqmod}). The system,
a pendulum with a harmonically modulated suspension point, exhibits parametric resonance as shown by Mathieu for the choice of $\omega^2 (t)$ in (\ref{mathieufreq}). Write the Floquet operator of the system as a product of $N$ unitary operators for $n$ consecutive time intervals of length $T/N$,
\begin{equation}
{\cal F} 
   = {\cal T} \exp \left[ - \frac{i}{\hbar} \int_{0^-}^{T^-} dt \hat H (t) \right]    
   = \prod_{n=0}^{N-1} U(t_{n+1}, t_n) \, , \qquad t_0 = 0^-, \, t_N = T^- \, .
\label{decompose}
\end{equation}
In the limit $N\to \infty$ the length of the time intervals goes to zero, and one has approximately
\begin{eqnarray}
U(t_{n+1}, t_n) 
       &=& {\cal T} \exp \left[ - \frac{i}{\hbar} \int_{t_{n}}^{t_{n+1}} dt 
         \left( \frac{{\hat p}^{2}}{2m} + \frac{m \omega^{2}(t)}{2}{\hat x}^{2} \right) \right] \\ \nonumber
       &\simeq& \exp \left[ - \frac{i}{\hbar} \left( \frac{{\hat p}^{2}}{2m} 
                     + \frac{m \omega_n^{2}}{2}{\hat x}^{2} \right)\frac{T}{N}  \right] \, ,
\label{approx}
\end{eqnarray}
where the number $\omega_n^2$ takes a value between $\omega^2(t_{n+1})$ and $\omega^2(t_n)$.
Being quadratic in position and momentum, any two adjacent exponentials can be entangled by means of a Baker-Campbell-Hausdorff relation. The result is another exponential 
bilinear in $\hat x$ and $\hat p$ as follows from the algebraic properties (\ref{splitalgebra}). Repeating this process for ever larger values of $N$, the Floquet operator tends to  
\begin{equation}
{\cal F} = \exp \left[ - \frac{i}{\hbar} \left (u\hat p^{2}+v\hat x^{2} 
           + \frac{w}{2}(\hat x\hat p + \hat p\hat x \right) \right]  
            = \exp \left[ - \frac{i}{\hbar} \hat  H_{\mbox{\footnotesize eff}}^\omega T \right]  \, .
\label{finalflo}
\end{equation}
According to the values of the parameters $u$, $v$, and $w$, the effective Hamiltonian $\hat  H_{\mbox{\footnotesize eff}}^\omega $ necessarily will be (unitarily equivalent to) one of the three possible 
types discussed exhaustively in Section 5. As long as the frequency modulation is periodic, no qualitatively different behaviour will occur within this class of systems. Clearly, the separation of the parameter space into stable and unstable regions will depend subtly on the actual function $\omega^2(t)$. 
%
\section{Discussion and outlook}
\label{disc}
The main result of the present paper is the insight that the concept of classical parametric resonance has a well-defined quantum analogon.
Classically, in a periodically driven linear system three qualitatively different types of motion are possible, according to the chosen parameter values. This division of parameter space is mirrored quantum mechanically by qualitatively different spectra of the Floquet operator. Its eigenfunctions associated with a classically stable
region are normalizable, and they become singular when moving into the classically instable parameter region.  This has been made explicit 
by constructing them in terms of (improper) position eigenstates. 

The class of time-dependent harmonic oscillators is well-known to provide insight into various aspects of classical and quantum mechanics. Therefore, some of the results can be found implicitly in work dealing with other properties of such systems. Closest to the present approach is, maybe, work by Perelomov \cite{perelomov86}: in a discussion of coherent states, a group-theoretical approach to driven linear systems is presented based on work in \cite{popov+69,popov+70}. Further, the paper \cite{facchi+00} deals with parametric resonance and its generation  within the realm of quantum optics. In this context, \cite{ferrari98} is also interesting, where exponential divergence of the energy expectation value has been derived for an oscillator with periodically modulated mass--- however, no global picture of quantum parametric resonance has been established. 

The impact of an instaneous mass change of a quantum harmonic oscillator also gives rise to squeezing, and the impact on the variance of position, for example, has been studied in \cite{janszky+93}. 
Non-periodic frequency modulations have been studied by various authors, starting with the now well-known Caldirola-Kanai oscillator with exponentially increasing mass \cite{caldirola41,kanai48} which, effectively, gives rise to damped motion of the oscillator. Among other work, exact solutions have been found for a polynomial 
time-dependence, $\omega^2 (t) = \omega^2 t^b, b>0$ in \cite{kim94}.    

Finally, it is worthwhile to point out that the periodically driven oscillator systems studied in the present paper suggest a natural generalization: modify the amplitude of the kick according to 
\begin{equation}
\hat {H}(t)
    = \hat {H}_0 + 
       \frac{\alpha}{2} \left( V(\hat{x})\hat{p}
      + \hat{p}V(\hat{x}) \right) \delta_T(t) \, ,
\label{qmoscillwithV}
\end{equation}
where $V(x)$ is some smooth function. Such a kick can be shown to  
generate {\em nonlinear} maps of configuration space onto itself \cite{morath96}. It will be shown elsewhere that features such as quasiperiodicity and a devil's staircase, which are characteristic of classically chaotic dynamics, also play a role in the time evolution of such a quantum system.
\end{document}